\batchmode
\makeatletter
\def\input@path{{\string"C:/Users/Stanislaw/Desktop/Paper PRL/\string"/}}
\makeatother
\documentclass[twocolumn,english]{revtex4-1}
\usepackage[LGR,T1]{fontenc}
\usepackage[latin9]{inputenc}
\usepackage{color}
\usepackage{graphicx}
\usepackage{setspace}
\usepackage{subscript}

\makeatletter

\DeclareRobustCommand{\greektext}{%
  \fontencoding{LGR}\selectfont\def\encodingdefault{LGR}}
\DeclareRobustCommand{\textgreek}[1]{\leavevmode{\greektext #1}}
\DeclareFontEncoding{LGR}{}{}
\DeclareTextSymbol{\~}{LGR}{126}

\makeatother

\usepackage{babel}
\begin{document}

\title{Interplay between magnetism and sodium vacancy ordering in Na\textsubscript{x}
CoO\textsubscript{2}}

\author{S. Galeski\textsuperscript{{*}}\textsuperscript{}, K. Mattenberger,
B. Batlogg}

\address{Solid State Physics Laboratory, ETH Zurich, CH-8093 Zurich, Switzerland }

\email{Corresponding author: galeskis@phys.ethz.ch}

\selectlanguage{english}%
\begin{abstract}
Using a combination of low temperature nano-calorimetry and X-ray
diffraction we identify three temperature regimes characterized by
distinct Na ordering patterns (low temperature up to 290~K, intermediate
290-340~K and high above 340~K). Through freezing-in of these patterns
we established the two key roles sodium intercalation plays in the
formation of the magnetic ground state: supplying the proper electron
count for in-plane ferromagnetic interaction and through the 3D sodium
ordering providing the inter-plane antiferromagnetic exchange path.
\end{abstract}
\maketitle
When first synthesized Na\textsubscript{x}CoO\textsubscript{2} was
considered a good candidate material for manufacturing rechargeable
battery electrodes, yet it soon became clear that due to its higher
ion mobility Li\textsubscript{x}CoO\textsubscript{2} would be the
preferred material in energy storage. However because of the low abundance
of lithium in earth crust and the high costs of its extraction Na\textsubscript{x}CoO\textsubscript{2}
has again attracted interest as a simplest possible alternative.

While Na acts as a charge reservoir in Na\textsubscript{x}CoO\textsubscript{2}
battery electrodes its role in defining the complex physical properties
immanent to the CoO\textsubscript{2}layers is much more subtle. Na
tends to order and thus provides a complex Coulomb potential landscape
acting upon the cobalt 3d orbitals. This together with the highly
correlated character of the electrons in the CoO\textsubscript{2 }layers
leads to the emergence of a plethora of collective electronic ground
states, including charge, spin density waves and even superconductivity
when hydrated \citep{Helme2006,Kuroki2007,Mazin2005,Mendels2005,Schaak2003,Takada2003}.
Moreover even in the regions of the phase diagram without any electronic
ordering the properties of Na\textsubscript{x}CoO\textsubscript{2}
can be very unusual. This is especially visible in the Curie-Weiss
metal phase occurring for Na content just above the Lifshithz transition
\citep{YoshizumiDaisuke2007,Wang2003}.

A central question in the discussion of Na\textsubscript{x}CoO\textsubscript{2 }is
weather all these Fermi surface instabilities originate from the change
of electron count only or whether, for a given value of x, the geometric
arrangement of Na ions plays an essential role for electronic ordering
in the CoO\textsubscript{2 }layers \citep{Roger2007,Shu2009,Kuroki2007a}.

A first hint comes from the comparison of phase diagrams of Na\textsubscript{x}CoO\textsubscript{2}
with its sister compound Li\textsubscript{x}CoO\textsubscript{2},
the widely used electrode material due to its high Li mobility. Interestingly
for the same electron count, the ground states are found to be different
\citep{Motohashi2009}. One possible reason for this difference could
be the reduced ion mobility in Na\textsubscript{x}CoO\textsubscript{2}with
the unique advantage of being slow enough to be accessible to experiments
\citep{Carlier2009,Medarde2013}.

\textcolor{black}{Another surprising phenomenon is that for certain
sodium contents the low temperature ordering is dependent on cooling
history of the sample \citep{Schulze2008}. This raises the question
about the cause of the history effect: is it freezing-in of the high
temperature crystallographic structure or perhaps of the interlayer
sodium superstructures. }

\textcolor{black}{In this study we show, by using a combination of
calorimetry and X-ray diffraction, not only that the sodium superstructures
are responsible for the thermal history effect but also which sodium
superstructures induce particular magnetic transitions. Moreover by
rapid cooling (over 6000K/s) we were able to freeze-in the most disordered
sodium configuration proving that even short range sodium correlations
are sufficient to sustain low temperature magnetic order. }
\begin{figure}[h]
\textcolor{black}{\includegraphics[scale=0.65]{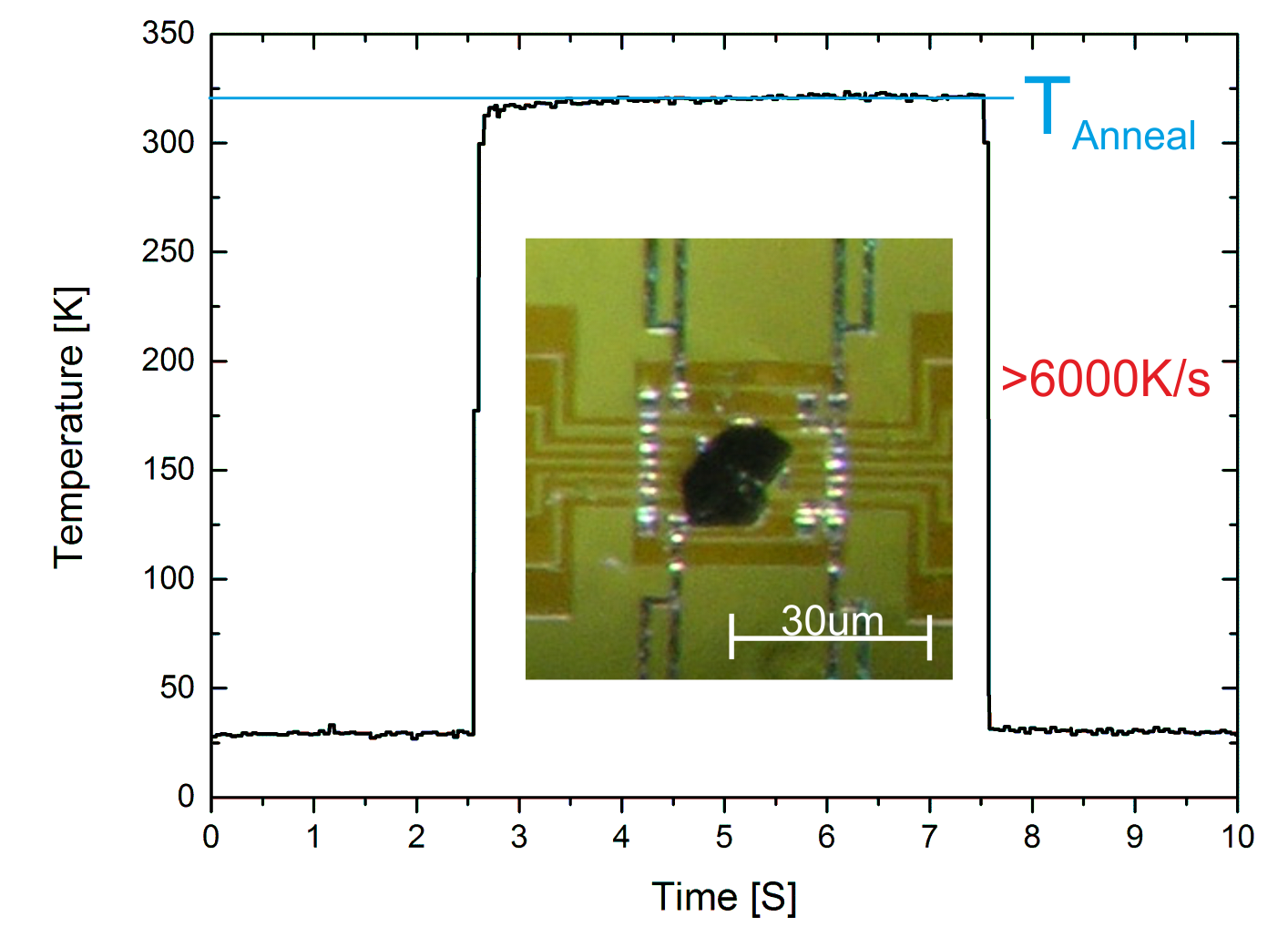}}

\textcolor{black}{\caption{\label{fig:A-typical-temperature}A typical temperature profile to
freeze-in the Na configuration established at the annealing temperature,
by subsequent quenching at a rate faster than 6000 K/sec. }
}
\end{figure}

\textcolor{black}{The difficulty in comparing the low temperature
magnetic order with sodium superstructures lies in the fact that they
can change upon cooling through the structural transitions of the
underlying lattices making the magnetic structures that could rise
due to high temperature sodium ordering inaccessible in standard experiments
\citep{Schulze2008}. To overcome this difficulty we have employed
a membrane calorimeter (Xensor Inc. \citep{Herwaarden}). With this
setup samples as small as 10-100~ng could be measured and more importantly
the temperature could be varied with heating and cooling rates up
to 6000~K/s enabling us to effectively freeze-in the high temperature
Na structures. Such fast cooling rates were achieved by keeping the
cryostat at 30~K and then passing a current pulse through one of
the resistors on the sample stage. A typical thermal cycle curve is
shown in Figure~\ref{fig:A-typical-temperature}. }

\textcolor{black}{The calorimetric measurements were performed using
the 3\textgreek{w} method \citep{Sullivan1968,Kohama2010} and were
cross-checked with standard measurements on large crystals from the
same batch performed on the Quantum Design PPMS specific heat option.
The crystals were grown by the floating zone method \citep{Prabhakaran2004},}\textcolor{black}{{}
the sodium content was determined using the known relation between
c-axis length and sodium concentration \citep{Schulze2008,Berthelot2011,Viciu2006}.}\textcolor{black}{{}
The crystal structure and the sodium super lattices were characterized
using the in-house Bruker Apex SMART II single crystal diffractometer
equipped with an Oxford Instruments temperature control system.}

\textcolor{black}{In the following we present results of three experiments
showing the relation between low temperature magnetic ordering and
high temperature sodium ordering. In the first section quenching the
sample from different high temperatures is shown to yield different
low temperature states. This is followed by a short discussion of
sodium dynamics and how different sodium structures evolve from one
into the other. As the last result we show X-ray Laue diffraction
patterns directly relating different low temperature magnetic states
to certain high temperature sodium states.}

\begin{figure}[!h]
\textcolor{black}{\includegraphics[scale=0.35]{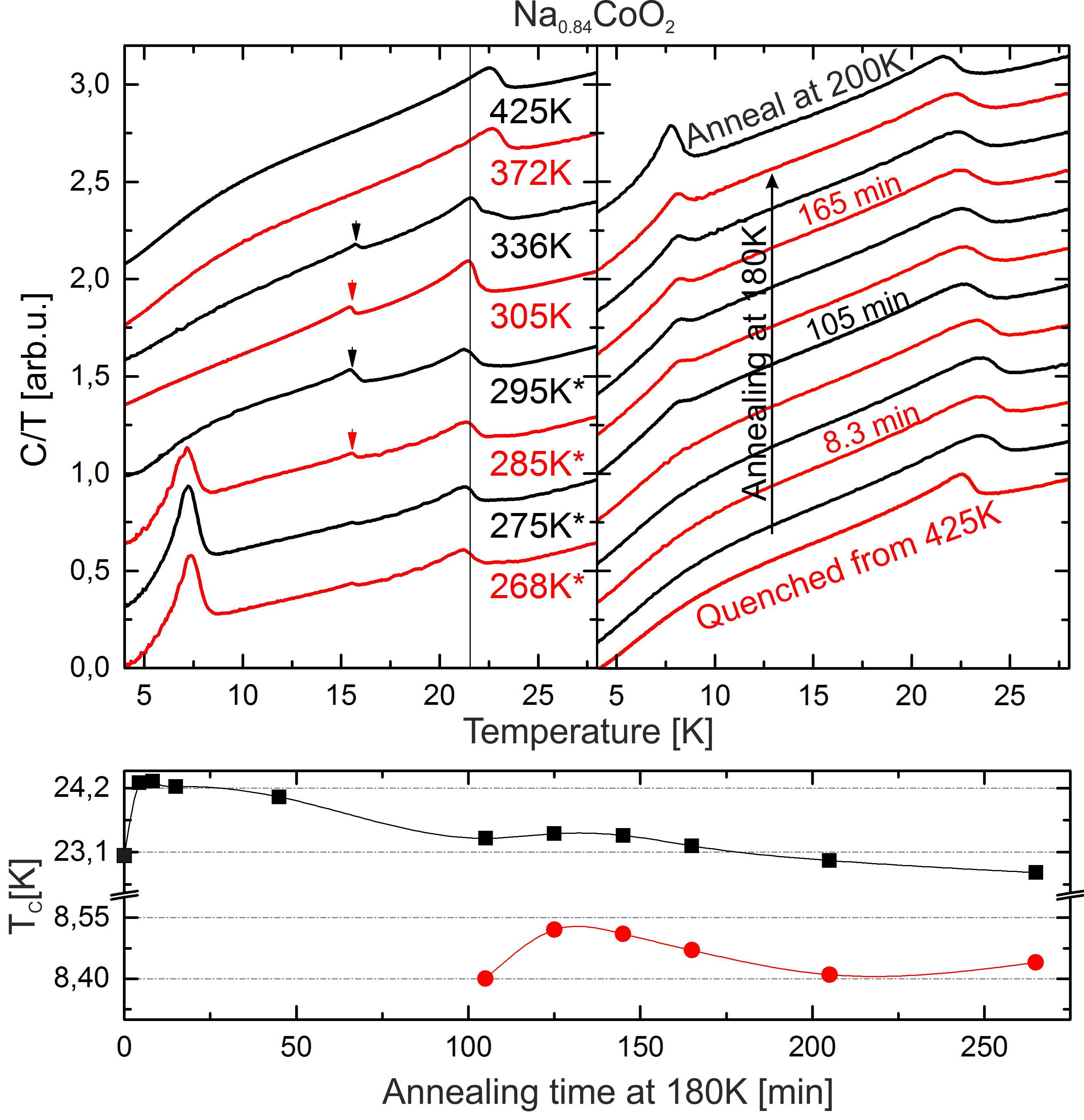}}

\textcolor{black}{\caption{\textbf{\textcolor{black}{\label{fig:Specyfic-heat-hata}}}\textcolor{black}{Emergence
of low temperature magnetic state for two different quenching and
annealing protocols (see text):}\textbf{\textcolor{black}{{} Left panel}}:Specific
heat data taken after annealing the samples at indicated temperatures
for 5s (the star indicates data from the second crystal, due to the
extreme temperature gradients between the sample stage and the cryostat
the SiN membranes occasionally broke. However the presented results
were cross checked on four other crystals from two batches and turned
out to be fully reproducible.). \textbf{\textcolor{black}{Right panel}}:
Evolution of the magnetic order with the change of sodium structure
from most disordered to the low temperature ordering. The sample was
prepared by quenching from 425K and then subsequently annealed at
180K. \textbf{\textcolor{black}{Bottom panel:}}\textbf{\textcolor{red}{{}
}}\textcolor{black}{Transition temperature evolution as a function
of annealing time.}}
}
\end{figure}

\textcolor{black}{To establish the boundaries defining the relation
between magnetism and sodium ordering in Na\textsubscript{0.84} CoO\textsubscript{2}
we have performed a series of annealing experiments. The sample was
initially slowly cooled (1K/min) from 300~K to 30~K and then annealed
for 5 seconds at successively higher temperatures with subsequent
quenching back to 30~K for specific heat measurements. Thanks to
the fast cooling rates (over 6000K/s figure \ref{fig:A-typical-temperature})
a direct mapping was possible between sodium configurations characteristic
for particular elevated temperatures and the low temperature magnetic
state they would induce in the CoO\textsubscript{2} layers. Three
distinct magnetic phases can be induced in the same Na\textsubscript{\textcolor{black}{0.84}}CoO\textsubscript{\textcolor{black}{2}}
crystal by freezing-in different Na configurations (Figure \ref{fig:Specyfic-heat-hata},
left panel). When the sample is cooled down slowly or annealed at
temperatures between where the temperature activated sodium mobility
becomes visible on the scale of minutes and 286~K the magnetic state
is characterized by two specific heat peaks at 8K and 21.5K. On subsequent
annealing at slightly higher temperatures the 8~K peak diminishes
slowly and another begins to form at 15K. This change coincides with
the previously reported structural transition taking place at around
290~K, it separates the 'low temperature' and 'intermediate temperature
regions'.}

\textcolor{black}{The magnetic order associated with the sodium structure
above 336K, the 'high temperature' region, is yet different: all transitions
previously reported \citep{Schulze2008,Mendels2005} are suppressed
and replaced by a single specific heat anomaly at 23~K. We have annealed
the sample up to 770~K and found no further modification of the low
temperature magnetic state.}

\begin{figure}[!h]
\begin{spacing}{0}
\noindent \begin{centering}
\textcolor{black}{\includegraphics[scale=0.45]{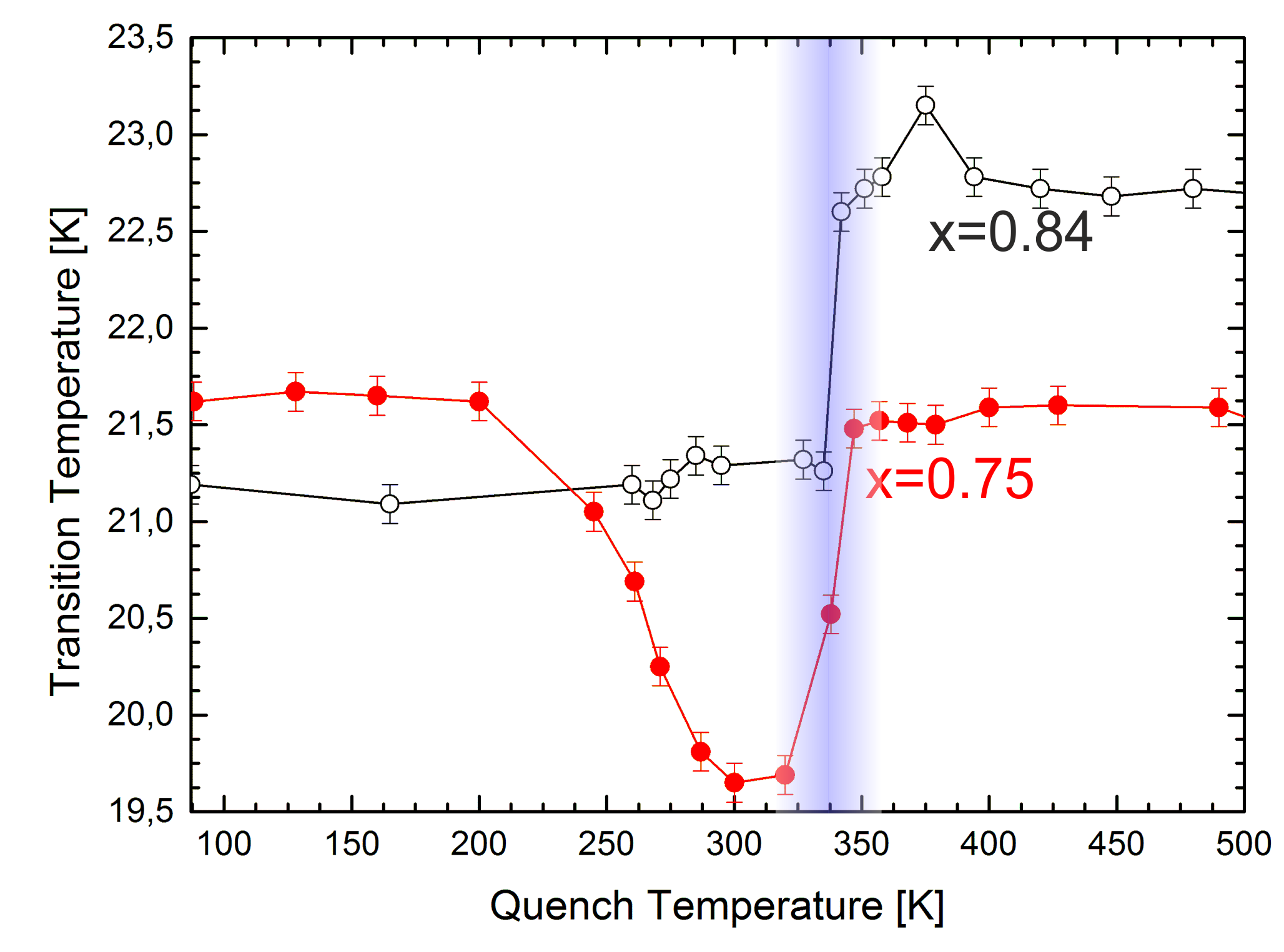}}
\par\end{centering}
\end{spacing}

\textcolor{black}{\caption{\label{fig:Variation-of-the}Variation of the transition temperature
of the highest temperature magnetic phase transition as a function
of quenching temperature for two Na concentrations. The blue region
indicates the location of the sodium lattice melting. }
}
\end{figure}

\textcolor{black}{In the same way we have also studied another composition
- Na\textsubscript{\textcolor{black}{0.75}}CoO\textsubscript{\textcolor{black}{2}}
and observed less spectacular changes in the low temperature magnetism:
a slight suppression of T\textsubscript{\textcolor{black}{N}} for
the sodium configuration in the 250-330~K range, figure \ref{fig:Variation-of-the}. }

\textcolor{black}{To better understand the dynamics and evolution
of sodium structures we have performed an annealing experiment in
which the sample was prepared in the high temperature state by quenching
it from 425~K and then annealed for short periods of time at 180K
with its low temperature specific heat probed between the annealing
steps. At 180~K Na is barely mobile enough to rearrange to the equilibrium
configuration after hours of annealing, so that transient magnetic
states can be monitored precisely (Figure~\ref{fig:Specyfic-heat-hata},
right panel). }

\textcolor{black}{}

\textcolor{black}{The most striking observation is that the 22~K
and the 23~K peaks seem to be continuously connected. Except for
the rapid change in the first 4 minutes of annealing, the shape of
the specific heat anomalies and the critical temperature vary very
slowly (figure~\ref{fig:Specyfic-heat-hata}, bottom panel). Surprisingly
this abrupt evolution, increased the transition temperature to 24~K.
One explanation of this phenomenon might be that in the first minutes
of annealing the sodium atoms just move to the higher symmetry position
between the Co atoms \citep{Huang2004} without moving far from the
position they were frozen-in. This way the exchange slightly strengthens
while maintaining the general high temperature sodium structure.}

\textcolor{black}{More interestingly the 8~K magnetic state evolves
from the 22~K in a direct way without the appearance of the 'intermediate'
15~K peak. Furthermore once the 8~K phase starts to emerge its transition
temperature evolves smoothly in direct correlation with the variations
of the peak at 22-24~K (figure~\ref{fig:Specyfic-heat-hata}, bottom
panel), suggesting that if there is any phase separation it will be
on a scale short enough for both phases to interact.}

\textcolor{black}{After having established the three temperature regions
associated with different magnetic order we searched for Na ordering
patterns performing XRD experiments at temperatures between 260~K
and 360~K on both x=0.75 and x=0.84 samples. Indeed we have found
distinct super lattice reflections that are directly linked to these
regions.}

\textcolor{black}{}

\textcolor{black}{}
\begin{figure}[!h]
\begin{centering}
\textcolor{black}{\includegraphics[scale=0.45]{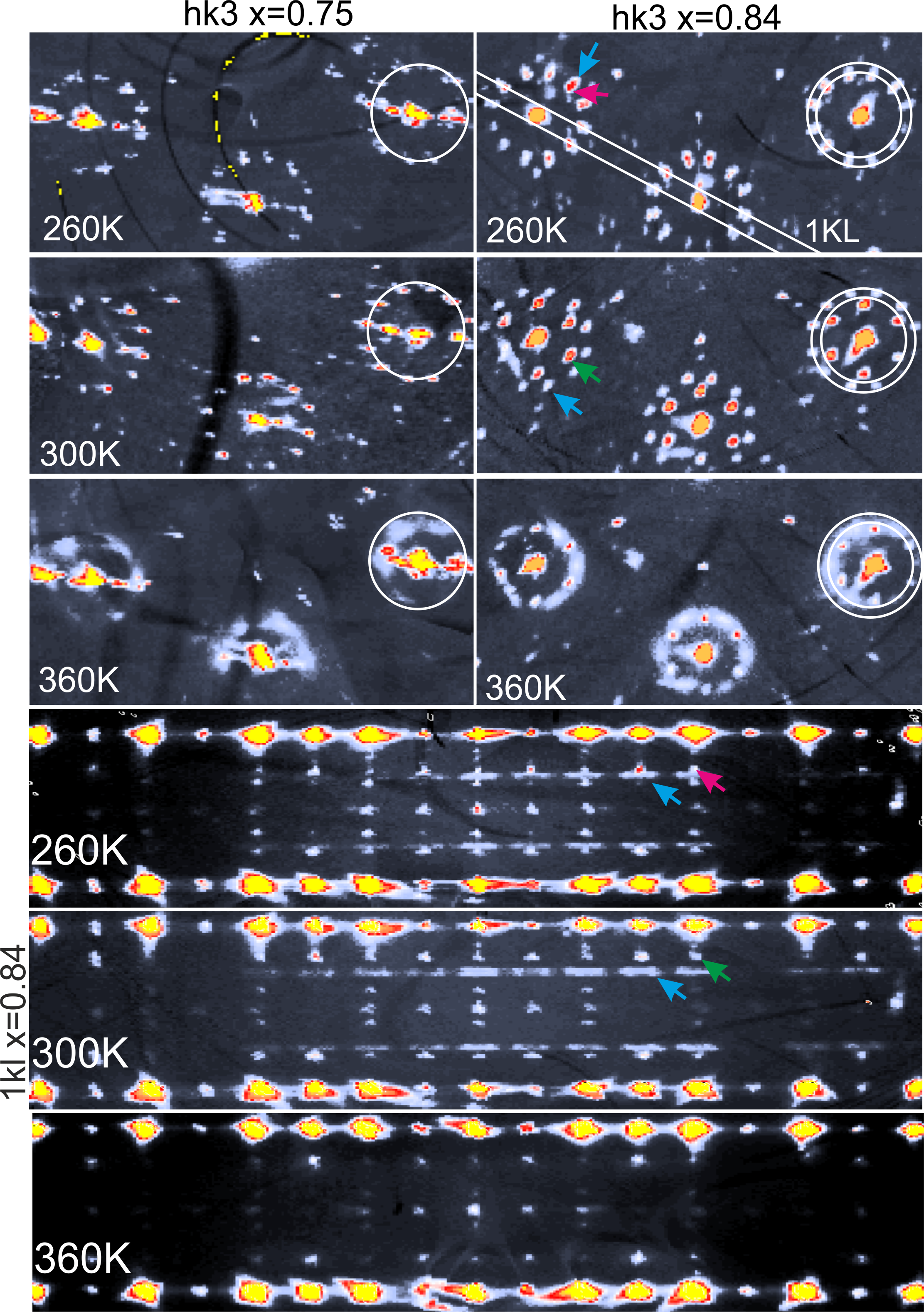}}
\par\end{centering}

\textcolor{black}{\caption{\label{fig:X-ray-diffraction}In and out of plane Na superstructures:
X-ray diffraction patterns showing sodium superstructures characteristic
for different temperatures. The images are hk3 planes measured on
a single crystal Na\protect\textsubscript{0.84} CoO\protect\textsubscript{2}(right
column) and Na\protect\textsubscript{0.75} CoO\protect\textsubscript{2}(left
column). The white circles are a guide to the eye to visualize the
change in modulation length.\textcolor{black}{{} The white lines in
the top right image indicate the integration range used for obtaining
the 1kl cuts. The coloured arrows indicate reflections of distinct
superstructures. The colour-code is maintained throughout paper ($\sqrt{13}$a
- blue, $\sqrt{19}$a - purple, 5a - green).}}
}
\end{figure}

\textcolor{black}{For x=0.84 below 290K the Bragg peaks belonging
to the CoO\textsubscript{2} lattice are surrounded by superstructure
reflections which form two 'rings' of twined superstructures, as indicated
by in the upper row of figure~\ref{fig:X-ray-diffraction}. The superstructures
can be indexed as tri- and di- vacancies (x=11/13=0.8462 and x=16/19=0.8421
respectively), on a hexagonal lattice with unit vectors $\sqrt{19}a$
and $\sqrt{13}a$ rotated by 2.5 and 15.9 degree with respect to the
original unit cell vectors. On crossing the 290~K structural phase
transition the twin $\sqrt{19}a$ reflections merge into a 'hexagon
of hexagons' pattern \citep{Morris2009}.}

\textcolor{black}{}
\begin{figure}[t]
\begin{centering}
\textcolor{black}{\includegraphics{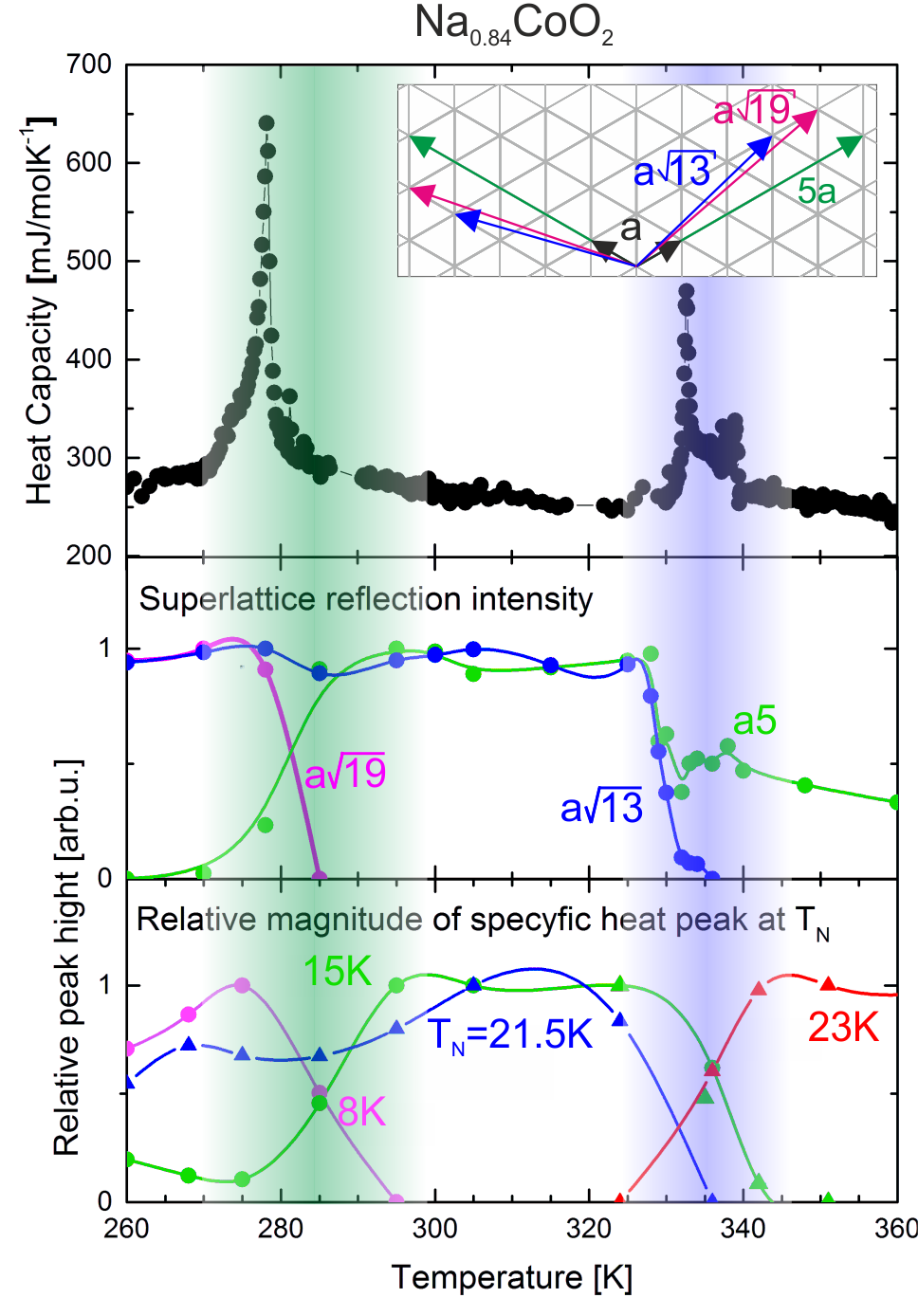}}
\par\end{centering}

\textcolor{black}{\caption{\label{fig:Comparison-of}Summary linking calorimetric and structural
data measured on Na\protect\textsubscript{0.84}CoO\protect\textsubscript{2}.
\textbf{\textcolor{black}{Top panel: }}\textcolor{black}{High temperature
heat capacity. }\textbf{\textcolor{black}{Middle panel:}}\textcolor{black}{{}
Normalized intensities of superstructure reflections taken from the
images of the right panel, for the Laue diffraction patterns see supplement..
}\textbf{\textcolor{black}{Bottom panel:}}\textcolor{black}{{} Relative
magnitude of specific heat anomalies. }}
}
\end{figure}

\textcolor{black}{Even more pronounced changes occur above 335~K
- a nearly complete disappearance of superstructure Bragg peaks. Although
there seems to be a reminiscence of the hexagon of hexagons patten
its intensity is suppressed (figure \ref{fig:Comparison-of} right
third panel from the top). Instead of a series of sharp Bragg peaks
a blurred halo develops around the main reflections. This is an indication
of the melting of the sodium superstructure lattice, and a formation
of what could be called a correlated sodium vacancy liquid: A state
in which the sodium vacancy droplets are still present due to Coulomb
repulsion but without angular ordering on the scale probed in an XRD
experiment. At the same time the average sodium structure modulation
length increases as best seen by comparison with the white circles.}

\textcolor{black}{An inspection of the 1kl cuts of the reciprocal
space reveals more details about the superstructures: Well defined
sodium ordering is present also along the c-axis. In particular the
tri-vacancy layers follow the periodicity of the crystal (red and
green arrows). On the other hand the location of the di-vacancy peaks
suggests an additional modulation tripling the original unit cell
height in accordance with previous findings (blue arrows)\citep{Shu2009}. }

\textcolor{black}{Sodium superstructures for x=0.75 are slightly different:
there is no detectable change on crossing 290~K, possibly due to
resolution limitations of our setup: the crystallographic transition
at 290K is very subtle \citep{Medarde2013}, thus the possible superstructure
modification is expected to be miniscule. However melting of the sodium
superstructures above is 335~K still present. }

\textcolor{black}{Figure~5 summarizes the salient features of the
structure and heat capacity studies as a function temperature:reflection
intensities for various super-lattice peaks, specific heat peaks heights
for various magnetic orderings and the specific heat anomalies that
mark the boundaries between three discussed temperature regions.}

\textcolor{black}{The specific heat anomaly around 280~K coincides
with the known structural transition of the unit cell from orthorhombic
to monoclinic. At this point it is not clear weather the 290~K peak
is associated directly to the CoO\textsubscript{\textcolor{black}{2}}
lattice transition or with the sodium rearrangement. }

\textcolor{black}{No structural transitions of the CoO\textsubscript{\textcolor{black}{2}}
units were reported for the 335~K region \citep{Medarde2013}, however,
neutron powder diffraction studies suggested a movement of Na ions
away from the most symmetric position in the unit Co - triangles \citep{Huang2004}.
This supports the proposed emergence of a sodium vacancy liquid.}

\textcolor{black}{The central result of this study is shown in figure~\ref{fig:Comparison-of}:
the super-lattice peak intensities and the specific heat anomalies,
signatures of magnetic states are closely correlated. In particular
the 8~K transition is related to the $\sqrt{19}a$ superstructure
which disappears at 290K replaced by a 5a superstructure pattern responsible
for the magnetic order at 15~K.}

\textcolor{black}{This pattern as suggested by Morris et al. \citep{Morris2009}
could indicate a formation of disordered sodium vacancy stripes. These
stripes would in a natural way lead to the modification of the effective
dimensionality of the magnetic interactions, thus explaining the observed
pronounced magnetic fluctuation tail preceding the 15~K transition
\citep{Kanter2015}. }

\textcolor{black}{On cooling the 290~K transition involves a locking
in of the modulation direction to the prevalent $\sqrt{13}a$ modulation.
In order to stay commensurate with the CoO\textsubscript{\textcolor{black}{2}}
lattice the modulation length slightly decreases from 5a to $\sqrt{19}a$
(inset of figure~\ref{fig:Comparison-of}).}

\textcolor{black}{The 335~K transition is of a different kind: there
is no lattice structural transition and no transition between different
superstructures. On crossing the transition temperature superstructure
Bragg peaks fade away. Freezing-in of this disordered state could
be pictured as creating a sodium vacancy glass and is associated with
the disappearance of all so far known magnetically ordered phases
and an emergence of a new state with transition temperature of 23~K
and with a spin flop transition \citep{Helme2006} at 12~T (supplementary
material). Suggesting that the presence of sodium long range order
is not necessary for magnetism to develop.}

\textcolor{black}{Our studies show that Na\textsubscript{0.84\textcolor{black}{{} }}CoO\textsubscript{2\textcolor{black}{{} }}has
a strong tendency towards A-type anti-ferromagnetism regardless of
the details of the sodium arrangement. Due to the long wavelength
of the sodium superstructure modulation and the overall hexagonal
symmetry of the sodium patterns the itinerant-electron ferromagnetism
in the CoO\textsubscript{2} layers would not be significantly altered
by modifications of sodium ordering. As was shown through measurements
of the magnetic contribution to specific heat, magnetic fluctuations
extend well beyond 50K and thus any known 3D ordering temperature
\citep{Zorkovska2005}. }\textcolor{black}{On the other hand it was
shown by Johannes et al. \citep{Johannes2004} that the strength of
the inter-plane interaction is very sensitive to the availability
of exchange paths involving sodium sp\textsuperscript{2} orbitals.
In particular the calculations predict the interaction strength to
be 35\% stronger if the sodium is placed directly under the Co atom
(Na(1) position) instead of the centre of the triangle made by the
Co atoms - (Na(2) position). }

\textcolor{black}{Our XRD results indicate that all the superstructures
are well ordered along the c-axis. This suggests that the reason for
the occurrence of different transition temperatures is not the modification
of the in-plane ferromagnetic fluctuations due to the Coulomb potential
or a change of the stacking order of vacancies of different nature
(di- tri- vacancies) but the modification of the inter-plane coupling
due to the change of relative location of vacancies in adjacent layers:
the $\sqrt{19}a$ and 5a superstructures are both tri- vacancy patterns
however they induce different T\textsubscript{\textcolor{black}{N}}'s. }

\textcolor{black}{In summary we have established the importance of 3D
sodium ordering as a key ingredient influencing the electronic properties
of Na\textsubscript{x}CoO\textsubscript{2}: (1) the sodium intercalation
provides the proper electron count sustaining the in-plane ferromagnetism
(2) it mediates the antiferromagnetic exchange between the layers.
This leads to a picture in which magnetism relies on the proper nesting
and electron count as suggested by calculations of Kuroaki et al.
\citep{Kuroki2007} but the exact value of T\textsubscript{N}is defined
by the nature of sodium ordering.}

Additionally we have shown that it is possible to freeze-in the disordered
sodium vacancy liquid arrangement. This could allow on one hand to
obtain though electrochemical reactions at elevated temperatures samples
of arbitrary sodium content and on the other to investigate which
phenomena are intrinsic to the CoO\textsubscript{2} layers and which
are emergent phenomena originating from the interplay of electronic
degrees of freedom with the sodium order. 
\begin{acknowledgments}
We would like to thank P.J.W. Moll, Y. Sassa and J.Kanter for their
input to this work and S. Gvasaliya for his assistance in performing
the XRD experiments.
\end{acknowledgments}

\bibliographystyle{apsrev4-1}
\bibliography{5C__Users_Stanislaw_Desktop_Paper_PRL_Biblio_1}

\end{document}